# Enhanced nonlinear optical response of single metal–dielectric nanocavities resonating in the near-infrared

Nicolò Maccaferri[1,¶], Attilio Zilli[2,¶], Tommi Isoniemi[3,4], Lavinia Ghirardini[2], Marzia Iarossi[4,5], Marco Finazzi[2], Michele Celebrano[2]*, and Francesco De Angelis[4]

[1]Department of Physics and Materials Science, University of Luxembourg, L-1511, Luxembourg, Luxembourg

[2]Department of Physics, Politecnico di Milano, I-20133, Milano, Italy

[3]Department of Physics and Astronomy, University of Sheffield, Sheffield S3 7RH, UK

[4]Istituto Italiano di Tecnologia, I-16163, Genova, Italy

[5]Dipartimento di Informatica, Bioingegneria, Robotica e Ingegneria dei Sistemi (DIBRIS), Università degli Studi di Genova, I-16126 Genova, Italy

*michele.celebrano@polimi.it

ABSTRACT

Harmonic generation mechanisms are of great interest in nanoscience and nanotechnology, since they allow generating visible light by using near-infrared radiation, which is particularly suitable for its endless applications in bio-nanophotonics and opto-electronics. In this context, multilayer metal-dielectric nanocavities are widely used for light confinement and waveguiding at the nanoscale. They exhibit intense and localized resonances that can be conveniently tuned in the near-infrared and are therefore ideal for enhancing nonlinear effects in this spectral range. In this work, we experimentally investigate the nonlinear optical response of multilayer metal–dielectric nanocavities. By engineering their absorption efficiency and exploiting their intrinsic interface-induced symmetry breaking, we achieve one order of magnitude higher second-harmonic generation efficiency compared to gold nanostructures featuring the same geometry and resonant behavior. In particular, while the third order nonlinear susceptibility is comparable with that of bulk Au, we estimate a second order nonlinear susceptibility of the order of 1 pm/V, which is comparable with that of typical nonlinear crystals. We envision that our system, which combines the advantages of both plasmonic and dielectric materials, might enable the realization of composite and multi-functional nano-systems for an efficient manipulation of nonlinear optical processes at the nanoscale.

INTRODUCTION

Multilayered metal-dielectric (MMD) cavities are widely used for light confinement and guiding at the nanoscale. So far, MMD nanocavities have successfully been implemented as negative index materials [1,2], planar waveguides [3,4], color filters [5], quantum yield enhancers [6–9], super-absorbers driving resonant gain singularities [10–12], as well as for hot-electron generation [13], manipulation of light scattering and absorption [14–18], ultrafast all-optical switching [19,20], and highly-sensitive detection of molecules [21]. The nonlinear optical (NLO) properties of these systems were theoretically addressed for their bulk form, for instance for second-harmonic generation (SHG) [22–24], and very recently also measured in continuous multilayered thin films [25]. Yet, to date, an experimental study on the NLO response of nanostructured MMD materials is still missing. In this direction, it was recently demonstrated that plasmonic nanostructures on-a-mirror configuration and supporting cavity gap plasmons can enable multiple and simultaneous nonlinear effects [26], a property that can be used, for instance, in multi-harmonic tissue imaging to increase selectivity [27]. In optics, SHG and third-harmonic generation (THG) mechanisms are of great interest, since they generate visible light from near-infrared (NIR) radiation, which is particularly suitable for applications in bio-photonics and opto-electronics. The three main reasons that make MMD nano-systems and the NIR frequency range ideal for both these applications are the following: (i) radiation with a wavelength longer than 750 nm falls within the transparency window of tissues and penetrates biological matter such as blood, brain/organs and skin [28,29]; (ii) dielectric materials like $SiO_2$ display an almost flat dispersion in the NIR range, allowing to design devices that introduce minimal pulse distortion [30,31]; (iii) Au nanostructures are biocompatible, easy to functionalize and, for a fundamental wavelength (FW) excitation in the NIR range, the absorption of the generated harmonics due to interband transitions is mitigated. Furthermore, it is well known that metal nanoparticles exhibit intense and localized field enhancements associated with plasmonic resonances, which can be conveniently tuned to the NIR, making them also ideal to enable nonlinear labelling in biological tissues, which in general display weaker SHG and THG signals, and enhanced nonlinear effects in opto-electronic devices [32].

Because of the sizeable values of $\chi^{(3)}$ for plasmonic metals in the visible range, the THG yield can be comparatively high in plasmonic nanoparticles at visible wavelengths. Nevertheless, the THG efficiency in Au is expected to significantly drop far from inter-band transitions, e.g. when the wavelength of the pump beam is tuned to the NIR [33–37]. However, it was recently demonstrated that THG can still be significantly high even when pumping in the NIR because of second-order cascade effects [38]. Unfortunately, the cubic symmetry of the Au lattice would forbid even-order nonlinear processes, such as SHG, to take place. This limitation has been recently alleviated by engineering specific nanoantennas featuring a broken-symmetry geometry, which enables the efficient exploitation of the high nonlinear currents developing at their surface **[39─43]**.

In this work, we address the nonlinear optical properties of metal-dielectric hybrid nano-systems. The investigated platform, which can also be realized using either common chemical synthesis [44─46] or bottom-up approaches [47─49], can enhance the NLO response of the nanoantenna by engineering the absorption and scattering processes **[16]** as well as exploiting the interface-induced local symmetry breaking typical of metal–dielectric multi-layers. We show that MMD nanocavities provide one order of magnitude enhancement of the SHG response compared to pure Au nanoresonators with the same geometry and displaying plasmonic resonances in the same spectral range. We also compare their SHG and THG performance when the axial symmetry is removed with a lateral cut, realizing structures similar to the ones described in Ref. **[43]**, and show that, in our experiments, such symmetry breaking play a minor role compared to the pivotal effect of the multilayered structure in enhancing NLO effects.

RESULTS AND DISCUSSION

MMD nanocavities made of 5 layers of Au [15 nm] alternating with 5 layers of $SiO_2$ [15 nm] were fabricated using gallium focused ion beam (FIB) milling (see **Methods** for more details) and designed to exhibit strong absorption in the NIR spectral range, by following the design rules presented in Ref. **[16]**. The nominal radius R of the structures spans from 100 nm to 250 nm in steps of 25 nm. Each structure was realized in several nominally identical replicas to check the reproducibility of the NLO properties of the single nanostructures (see **Figure 1a**). For comparison, we fabricated also bulk Au nanoresonators with the same amount of gold as in the MMD nanocavities (thickness 75 nm), and resonating in the same spectral range (see **Supplementary Figure S1**). For both MMD and bulk Au systems we also fabricated an additional set of nanostructures with a radial cut, which removes the axial symmetry. Such a cut induces a doubly resonant character in the scattering spectrum, which, along with the removal of the axial symmetry, leads to a strong field enhancement at both the FW and SH wavelengths and improves the outcoupling of the SH sources inside the nanostructure to the radiating modes **[41]**. **Figure 1a─c** show representative scanning electron microscopy (SEM) images of the MMD nanocavities with the radial cut. The nanostructures were obtained by milling a continuous multilayered film (or bulk gold) evaporated on a glass substrate and are encircled by a trench dug into the same film, which has been used as reference.

To assess separately the effect of multilayering and of symmetry removal, we collected SHG maps on both bulk Au nanoresonators and MMD nanocavities with axial symmetry (i.e. disc shape, from now on indicated with the symbol "O") and on structures with a cut. The latter were measured for different exciting polarizations, namely with the FW electric field either parallel ("∥" configuration) or perpendicular ("⊥" configuration) to the cut.

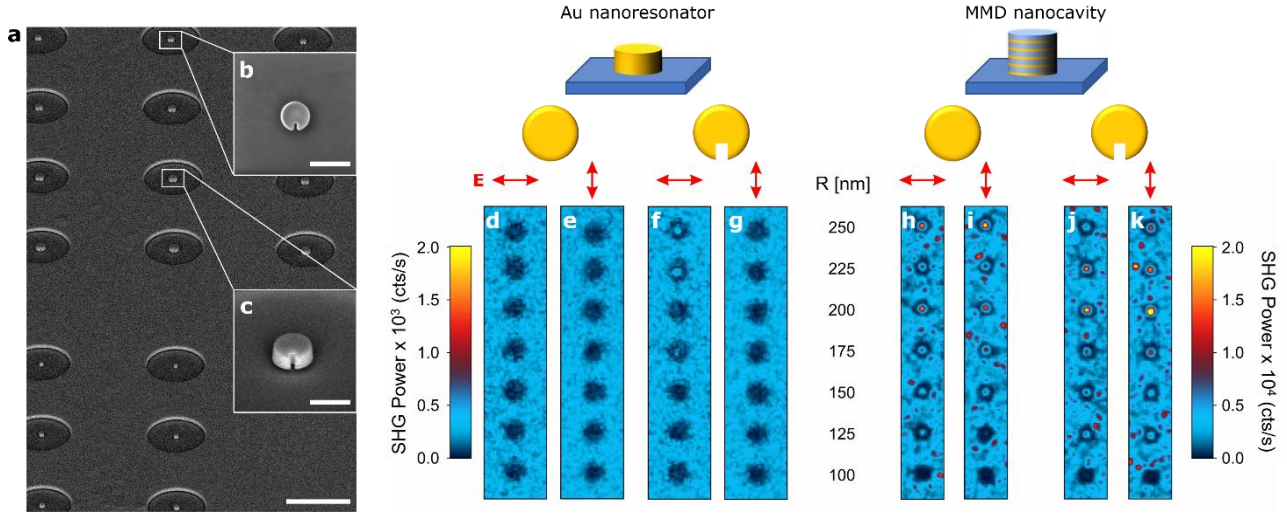

**Figure 1.** (a) Representative SEM images of the MMD nanocavities (scale bar 3 μm). Top (b) and side (c) views of MMD nanocavities with nominal radii of 175 nm and 225 nm, respectively (scale bar in both figures is 500 nm). (d–k) SHG confocal maps of Au nanoresonators (d–g) and MMD nanocavities (h–k) nanocavities using 500 μW average pump power (peak fluence 1 GW/cm$^2$). Notice that the color scales differ by one order of magnitude.

NLO measurements were performed on single nanostructures (see **Figure 1d─k**) by tightly focusing 500 μW average pump power, which is equivalent to a peak fluence of about 1 GW/cm$^2$, using a confocal microscope coupled with a pulsed laser at a telecom wavelength ($\lambda$ = 1554 nm). The setup, which is similar to that described in Ref. **[38]**, is illustrated in more detail in the **Methods** section and in **Supplementary Figure S2**. While the SHG collected from Au nanoresonators (**Fig. 1d─g**) is extremely weak, MMD nanocavities, both with (**Fig. 1j,k**) and without the radial cut (**Fig. 1h,i**), produce a steady SHG. In particular, in our experimental conditions, the "O" bulk Au structures do not yield any detectable SHG signal (**Figure 1d,e**). This is one more reason why we have also realized structures with a radial cut, which brings about a sizeable SHG even in the Au nanoresonators under "⊥" illumination (**Figure 1f**). On the other hand, the SHG signal of bulk Au cut discs is not detected with our instrumentation under "∥" illumination (**Figure 1g**).

For the MMD structures the behavior is different: the SHG yield increases significantly also for the centro-symmetric structures and a resonant behavior appears in the maps (see **Figures 1h–k**). We have estimated the SHG signal enhancement by evaluating the count rates in the maps collected on either the bulk Au or the MMD nanostructures with the radial cut under "⊥" illumination. At 500 µW average pump power (peak fluence 1 GW/cm$^2$), we measured a count rate of 0.6 kcts/s from the most efficient Au nanoresonator (the second from the top in **Figure 1f**), compared to 21 kcts/s from the most efficient MMD nanocavity (the third from the top in **Figure 1j**). This corresponds to an enhancement factor of about 35 for the MMD structures with respect to bulk Au structures. Under "∥" illumination the count rate for the most efficient MMD is 42 kcts/s. Although the assessment of the enhancement in this configuration is prevented by the unknown SHG signal on the bulk Au structures, by assuming the signal of the homogeneous Au nanoresonators to be of the order of the noise level, we can draw a lower bound value for the enhancement factor that is as high as 70. The same can be applied to the "O" geometry, which yields 15 kcts/s in the MMD case but no signal for the Au nanoresonators. Concurrently, the enhancement factor induced by the layering alone, which is obtained by comparing the SHG yield from the bulk Au and MMD film (0.5 and 6 kcts/s, respectively), is about 10. We thus ascribe the enhancement to the combination of three main mechanisms: (i) the appearance of a narrow absorption peak at the fundamental wavelength typical of these MMD nanocavities [**16**] (see also **Supplementary Figure S1**), (ii) the removal of the axial symmetry by the cut, and (iii) the local inversion symmetry-breaking at the metal-dielectric interfaces. Therefore, we conclude that shaping the MMD into resonant nanoparticles allows one to further boost the enhancement by over a factor 3 while considerably reducing the material fingerprint. The presence of the radial cut in the MMD structure does not provide a significant enhancement of the SHG with respect to multilayered "O"-shaped nanostructures. This is attributed to the Au–SiO$_2$ layered structure, which dominates over the effects induced by the axial symmetry breaking and the field enhancement in the cut.

By resolving the full nonlinear emission spectrum of the individual nanocavities of different radii in the visible range (see **Fig. 2a**) it is possible to observe, in addition to the strong SHG peak at 777 nm, a THG peak at 518 nm, as well as a sizeable and spectrally broad multiphoton photoluminescence (MPL) band typical of Au [**50, 51**]. The emission spectra in the "O" and "∥" cases exhibit similar features (see **Supplementary Figure S3**). To complete the characterization of the NLO behavior, we recorded the power dependence of SHG, THG and MPL from a resonant nanostructure ( radius = 200 nm, "⊥" case, **Figure 2b**) reported in **Figure 2a**, singling out each nonlinear emission signal by spectral filtering. The incident intensity on the sample, $I$, was varied by almost an order of magnitude, and the emitted power $P$ fitted by a power function $P = \text{cost} \cdot I^\alpha$. While the power coefficients $\alpha = 1.9$ and $\alpha = 2.8$ closely match the quadratic and cubic dependence expected for SHG and THG respectively, the power coefficient determined for the MPL is $\alpha = 3.0$, indicating a three-photon excited process [**52, 53**].

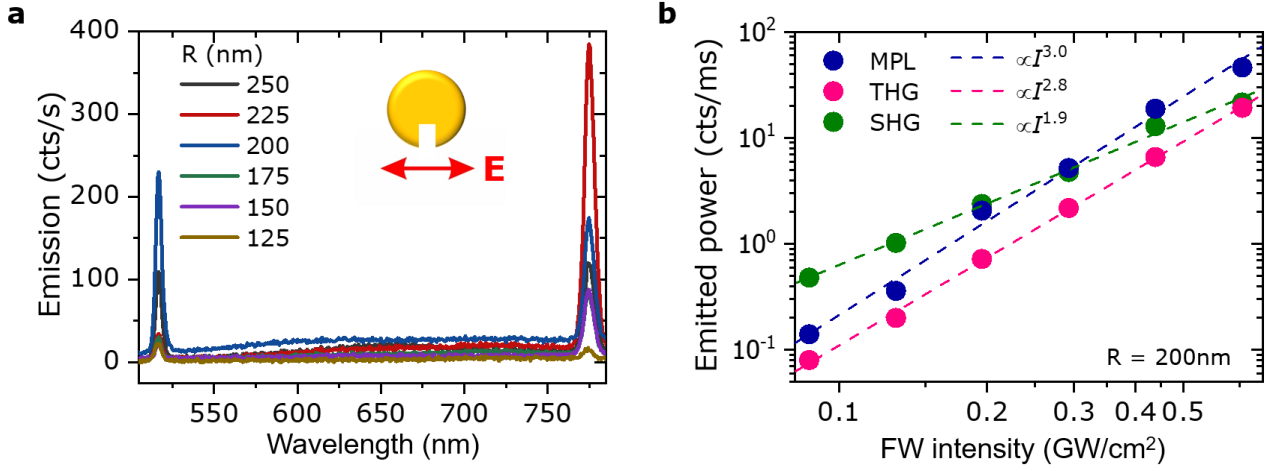

**Figure 2.** (a) NLO emission spectra by single cut structures of different radii as indicated in the legend. Excitation at the FW (1554 nm), "⊥" configuration. (b) NLO emission as a function of power by a single cut structure of radius $R = 200$ nm. The signal is optically filtered using narrowband bandpass filters (center/bandwidth in nm 775/25 for SHG, 520/40 for THG) and a combination of long-pass (cut-on at 550 nm) and short-pass (cut-off at 750 nm) filters for the MPL. The dashed lines are power fits to the data, with exponents indicated in the legend.

Considering the SHG and THG emission by the most efficient MMD nanocavity in the experiment (i.e. $R = 200$ nm) and accounting for the optical losses in the detection path to arrive to the overall emitted intensity, we experimentally estimate the following effective values for the second- and third-order nonlinear susceptibilities, $\chi^{(2)}_{eff}$ and $\chi^{(3)}_{eff}$, (see **Supplementary Note 1**): $\chi^{(3)}_{eff} \sim 10^{-20} m^2/V^2$, in line with THG measurements on Au in this wavelength range **[54]**, and $\chi^{(2)} \sim 1.5$ pm/V, which is comparable to some widely-employed second-order upconverting nonlinear crystals, such as potassium di-deuterium phosphate (KDP) **[55]**.

To gain further insight into the optical properties of the MMD nanocavities studied in this work, we simulated numerically their near-field response using the finite element method (see **Methods** for additional details). **Figure 3** shows the simulated enhancement of the electric field amplitude $|E/E_0|$ at the FW, SH, and TH wavelength by a single nanocavity with a 175 nm radius, which corresponds to the most efficient "O" structure. Following the nanostructure design, the largest enhancement is observed at the FW, whose field contributes quadratically to the SHG process and cubically to the THG one **[56]**. This enhancement is associated with the onset of an absorption peak, typical of these MMD systems at telecom wavelengths (see also **Supplementary Figure S1**). The field is mostly localized in the dielectric layers at the SH wavelength, and SHG is expected to occur mostly at the Au–SiO$_2$ interface, where the inversion symmetry of the two materials is broken. Conversely, at the TH wavelength the field is mostly confined in the metal layers, which have the highest $\chi^{(3)}$ and generate most of the TH emission.

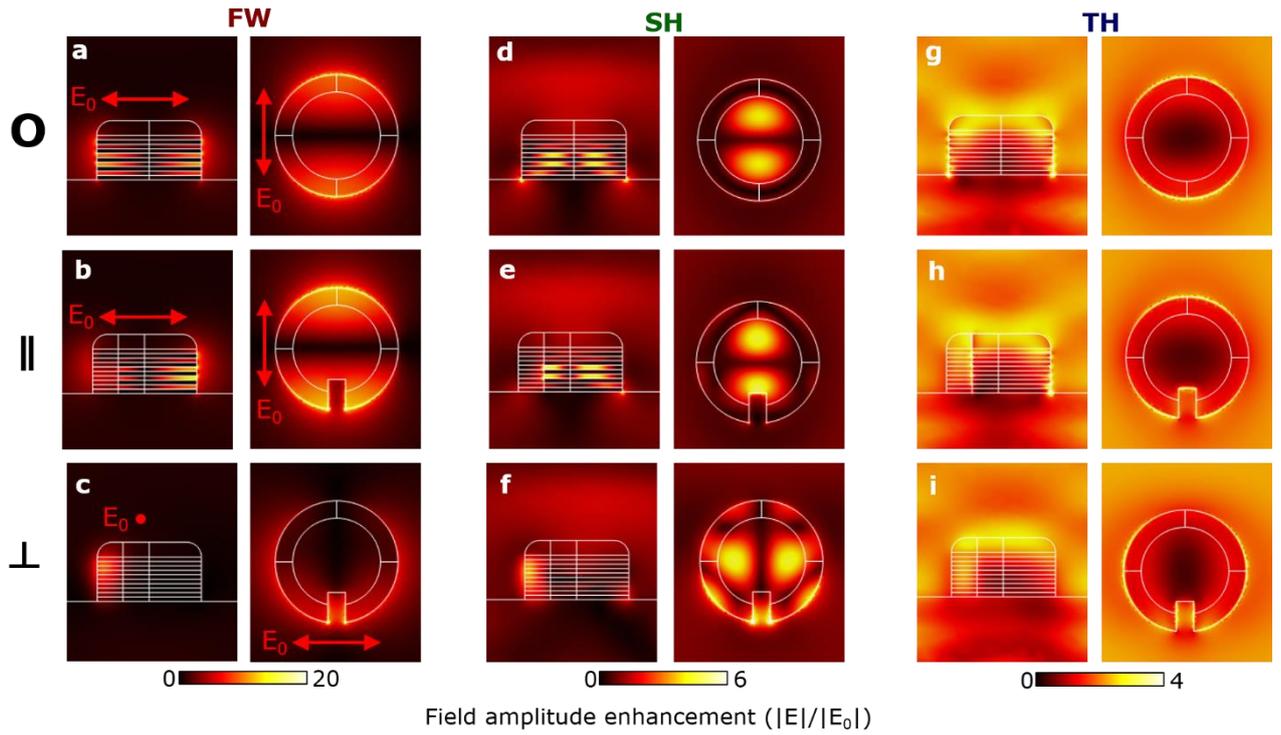

**Figure 3.** Simulated enhancement of the electric field amplitude $E$ with respect to the incident plane wave amplitude $E_0$ for a nanocavity with a radius of 175 nm. Panels (a), (b) and (c) are the computed near fields at the FW (1554 nm) for the O, ∥ and ⊥ cases. Panels (d), (e) and (f) are the computed near-fields at the SH (777 nm) for the "O", "∥" and "⊥" cases. Panels (g), (h) and (f) are the computed near fields at the TH (518 nm) for the "O", "∥" and "⊥" cases.

The spatial distribution of the mode excited at the FW is governed by a bonding dipolar resonance, as previously demonstrated in Ref. **[16]**. Notably, the presence of a cut inducing the in-plane symmetry breaking makes little difference with respect to the O case (**Figure 3a**) under "∥" illumination (**Figure 3b**). Apart from a slight variation of the field distribution, this also holds true at the SH and TH wavelengths for both "O" (**Figure 3d,g**) and "∥" (**Figure 3e,h**) cases. On the contrary, the symmetry breaking significantly modifies the spatial field distribution under "⊥" illumination at the FW (**Figure 3c**), SH (**Figure 3f**) and TH (**Figure 3i**) wavelength. In particular, strong light confinement occurs at the cut edges at the SH wavelength (**Figure 3f**). This analysis predicts a similar NLO response in the "O" and "∥" cases, while some differences are expected for the "⊥" case.

To evaluate the effect of the geometry, we performed a systematic study of the NLO response of the individual MMD nanocavities with radius increasing from 100 nm to 250 nm in steps of 25 nm. The measured SHG and THG powers, averaged over 10 nominally identical replicas of each structure, are reported in **Figure 4a** and **Figure 4b**, respectively, for all structures and excitation configurations. In Figures **4a,b**, the sizable error bars, representing the standard error of the mean, likely reflect the role of random defects introduced by the nanofabrication process, such as roughness in the thin films due to e-beam deposition and damage induced by the gallium beam.

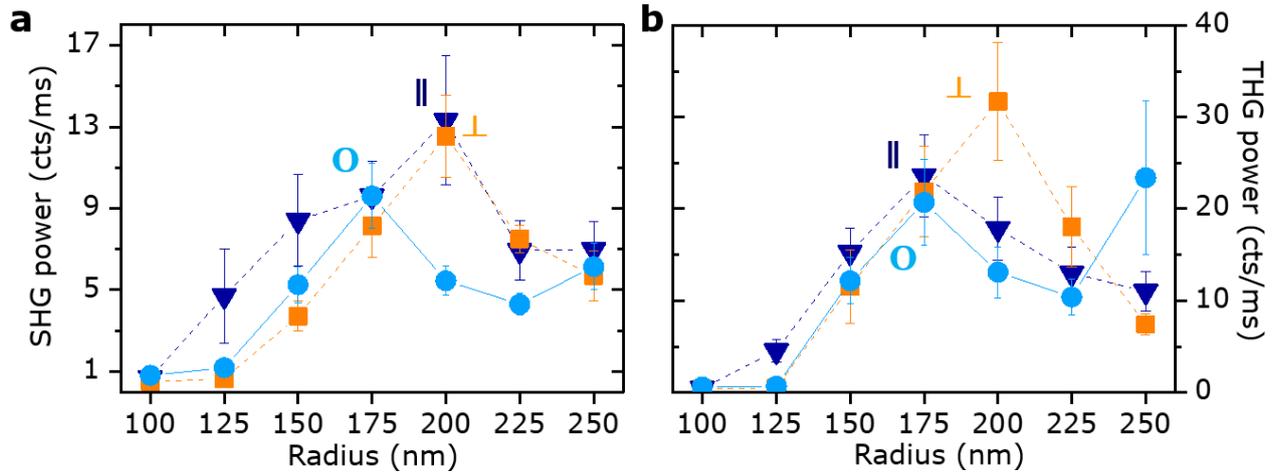

**Figure 4.** SHG (a) and THG (b) power (most intense pixel of the confocal image of the structure) of the MMD nanocavities. Each plotted point is the mean measured over 10 nominally identical replicas, and the error bars correspond to the standard error of the mean values. The signals are collected using an illumination average power of about 500 μW (peak fluence 1 GW/cm$^2$).

In the "O" structure, the NLO emission response is similar for both the SHG and THG with a maximum obtained for a radius of 175 nm, indicating that the emission is dominated by the resonant behavior at the FW. The cut nanostructures exhibit a more complex behavior. The SHG signal under both ∥ and ⊥ illumination features a maximum shifted towards longer radii (200 nm). Conversely, the THG peak for the ∥ case is found for the same radius (175 nm) as the "O" case, in agreement with the similar spatial distribution of the calculated near-field distributions in **Figure 3**. Conversely, the THG emission under "⊥" illumination peaks for the same radius (200 nm) that maximizes its SHG signal. The perfect match between the SHG peaks of the "∥" and "⊥" cases at R = 200 nm is a clear indication that SHG does not depend on the localized fields in the cut or on the consequent axial symmetry breaking. On the other hand, the shoulder in the SHG response for shorter radii in the ∥ configuration is unexpected and could be related to local random defects of the nanostructured MMDs.

To better assess the mechanisms underpinning SHG and THG, we numerically computed the scattering and absorption cross sections $\sigma_{sca}$ and $\sigma_{abs}$, at the FW, SH and TH wavelengths for all the cases nanostructures (**Figure 5a-f**). By inspecting the radius dependence of $\sigma_{sca}$ and $\sigma_{abs}$ at the FW, the "O" and "∥" cases (red curves in **Figure 5a,b** and **Figure 5c,d**, respectively) share a similar trend, whereas in the "⊥" case (red curves in **Figure 5e,f**) we observe a different behavior, as expected also from the near-field simulations. In particular both $\sigma_{abs}$ and $\sigma_{sca}$ peak at R = 225 nm, and this could explain the shift of both the SHG and THG peak efficiencies towards larger R with respect to the "O" case.

A simple approach to model the nonlinear emission from these structures, ultimately based on the Miller's rule **[57,58]**, is to combine the calculated radius-dependent linear cross sections (**Figure 5a-f**) into the nonlinear coefficients $\epsilon^{(2)} \equiv \sigma_{abs}(\omega)^2 \sigma_{sca}(2\omega)$ and $\epsilon^{(3)} \equiv \sigma_{abs}(\omega)^3 \sigma_{sca}(3\omega)$, assuming that the nonlinear process is governed by absorption at the FW wavelength, and by scattering at the harmonic emission wavelength (**Figure 5g,h**). These coefficients account for the coupling strength between the antenna and the radiation at the excitation and emission wavelengths. We emphasize that this description of the process does not consider the spatial overlap of the fields **[41]** at the two wavelengths (which can be qualitatively assessed from **Figure 3**) and the relative polarization. However, these aspects can be neglected in our case since the involved modes have a dominant dipolar character, and the nonlinear susceptibility tensors are diagonal. The measured nonlinear emission trends in **Figure 4a,b** are in qualitative agreement with the computed nonlinear coefficients (see **Figure 5g,h**); specifically, the coefficients reproduce the observed shift of the SHG and THG peaks towards larger radii introduced by the cut. In particular, by comparing **Figure 4b** with **Figure 5h**, it is evident that, despite small shifts and some broadening in the experimental results, the trend of $\epsilon^{(3)}$ mirrors the absorption resonances at the FW (see also **Figures 5a,c,e**). This trend confirms that this simple model grasps the main mechanism of THG in metal-based nanostructures. Conversely, the behavior at the SHG (**Figure 5g**) is less straightforward. The SH emission is again well-reproduced for the "O" case, with the higher SHG coinciding with the higher absorption at the FW (similar to THG). The main discrepancy is the position of the SHG peak for the "∥" case, which shifts to larger radii along with the "⊥" case in the experiment (Figure 4a) whereas it does not in the model (Figure 5g), where it coincides with the "O" case. We ascribe this difference to a mechanism not included in the model, namely the fact that for SHG the co-localization and co-polarization of the electric fields at the FW and SH becomes more relevant within the structure at the interfaces between the multilayers, rather than within the cut (see **Figure 3b,e**). This mechanism might be fostered by surface roughness or defects at the interfaces between layers, while the removal of the axial symmetry reduces the destructive interference of the SH radiation in the far field **[41]**. The perfect matching between the SHG efficiency in the "⊥" case and "∥" case for the 200 nm radius nanoparticle makes this geometry particularly interesting, since the SHG does not depend on the polarization of the impinging light.

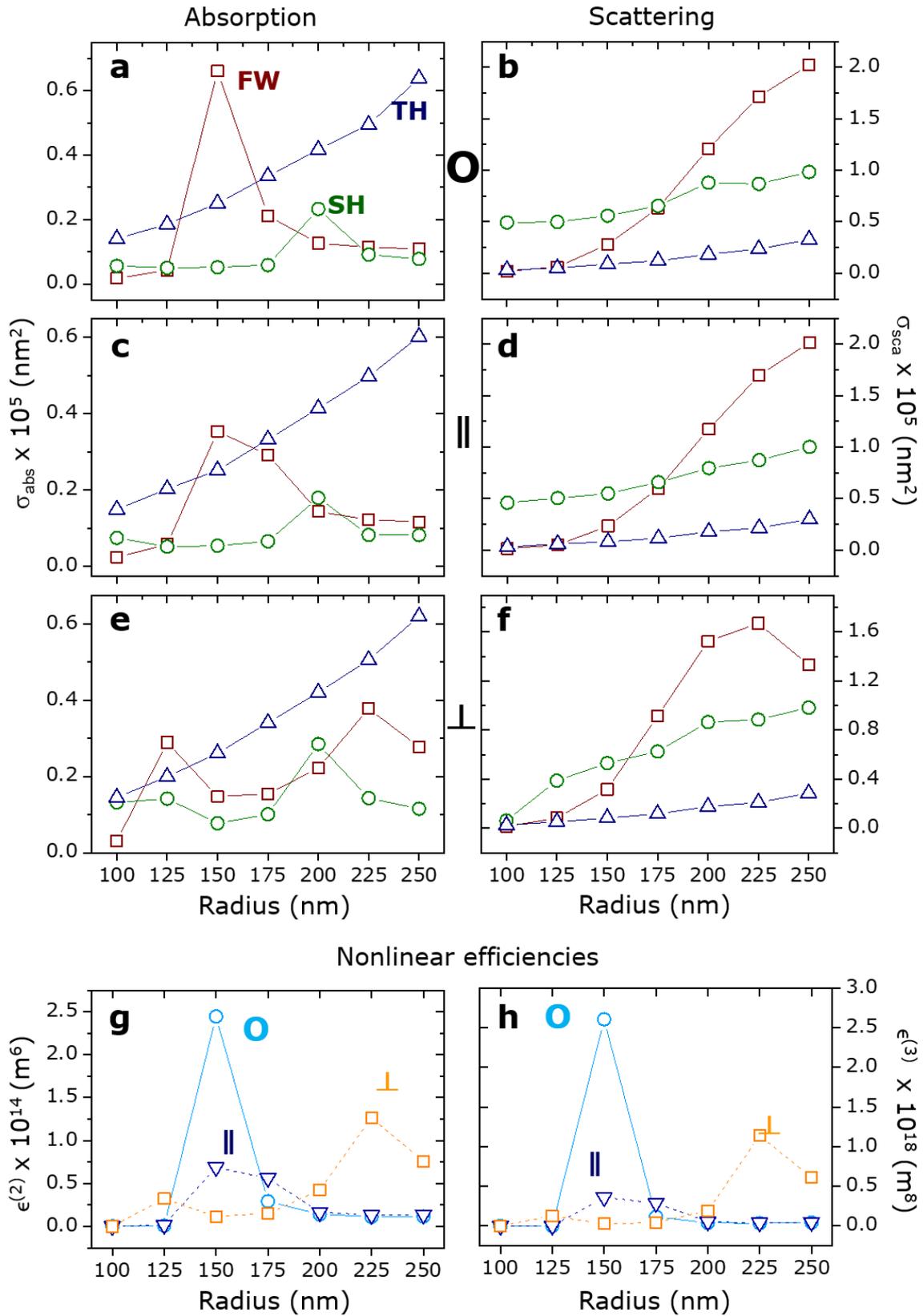

**Figure 5.** Calculated absorption (left) and scattering (right) cross-sections for the "O" (a,b), "∥" (c,d) and "⊥" (e,f) case, as a function of the structure radius at the FW (brown curves), SH (green curves) and TH (blue curves). Calculated (g) efficiencies $\epsilon^{(2)}$ and (h) $\epsilon^{(3)}$ as a function of the nanocavity radius.

CONCLUSION

We reported efficient second- and third-harmonic emission from multilayered metal-dielectric nanocavities, where the absorption at the fundamental wavelength is enhanced by tailoring the geometry of the structures. We proved that the presence of multiple metal-dielectric interfaces, rather than the breaking of the centro-symmetric geometry of the nanocavities, boosts the nonlinear optical generation. As a result, the nonlinear optical response is largely independent from the exciting polarization. Noticeably, we found that in our experimental conditions the second-harmonic emission is more than one order of magnitude larger than that of pure Au nanostructures with the same geometry and resonant behavior. We underline that these structures can be fabricated on a large scale by using hole-mask colloidal lithography also to achieve complicated structures [16,47-49,59], and detached from the substrate, thereby enabling their use, for instance, as nonlinear optical microscopy labels in live systems, where it is fundamental to control the emission of visible light from objects that resonate in the NIR spectral range and are located deep in the tissue. Currently, the NIR II transparency window (1000–1350 nm) is the subject of intense investigations and the III transparency window (1600–1870 nm) is emerging too thanks to its deeper penetration and lower photon scattering in various types of tissues, including brain. The choice of materials is important to provide high values of $\chi^{(2)}$ and $\chi^{(3)}$. We have found that our nonlinear susceptibilities are of the same order of magnitude of common nonlinear optical crystals, such as KDP. Other materials, such as lithium niobate ($LiNbO_3$), potassium dihydrogenphosphate ($KH_2PO_4$), barium titanate ($BaTiO_3$), bismuth ferrite ($BiFeO_3$), though having higher $\chi^{(2)}$ values, are not well-suited for *in-vivo* imaging due to their cytotoxicity [60,61]. On the contrary, Au nanoparticles and dielectric materials such as $SiO_2$ are widely used in nanomedicine thanks to their biocompatibility [62]. Finally, our approach is a first step towards merging the advantages of biocompatible plasmonic materials with those offered by the emerging field of soft polymers [63] by realizing composite and multi-functional nano-systems.

METHODS

**Sample fabrication.** The multilayers were deposited on a glass substrate via electron-beam evaporation (Kurt J. Lesker PVD75). The 1 mm-thick glass slide was cleaned with acetone, isopropanol, and oxygen plasma. The layers were deposited sequentially, starting with 15 nm of Au and continuing with 15 nm $SiO_2$ for five layers of each, and capped with an additional protective layer of 50 nm $SiO_2$. The pillar structures were created with focused ion beam milling using a gallium beam (FEI Helios NanoLab 650 dual beam system) at 24 pA, milling down to the glass surface around each pillar. The empty milled area around each structure was a circle 3 µm in diameter, and the resulting milling time was 6 min per structure. The pillars have a radius varying from 100 nm to 250 nm in steps of 25 nm, and the variant with cuts has nominally 50 nm-wide, half-radius long radial cuts. The control samples were fabricated by depositing 75 nm of Au, namely the same amount of the multilayers, followed by a top layer of 50 nm of SiO2 and then, realizing disks with FIB in a similar way as described for the MMD disks. The structures were imaged with scanning electron microscopy in the same system after milling.

**Numerical Simulations.** Calculations using the finite element method have been performed using Comsol Multiphysics, a commercial solver. The refractive index values of gold and SiO$_2$ have been taken from literature **[64,65]**. To simulate the optical properties of MMD nanocavities we have considered a simulation region where we specified the exciting ("background") electric field (a linearly polarized plane wave), and then we calculated the scattered field by a single nanocavity to extract observables such as the optical absorption and scattering cross-sections. The scattering cross-section is defined as $\sigma_{sca} = \frac{1}{I_0} \iint (\boldsymbol{n} \cdot \boldsymbol{S})\, dS$, where $I_0$ is the intensity of the incident light, $\boldsymbol{n}$ is the unit vector normal to the integration surface pointing outwards and $\boldsymbol{S}$ is the Poynting vector. The integral is taken over the closed surface of the meta-antenna. The absorption cross section equals $\sigma_{abs} = \frac{1}{I_0} \iiint Q\, dV$, where $Q$ is the resistive power loss density of the system and the integral is taken over the volume of the nanocavity.

**Nonlinear optical experiments.** The nonlinear optical measurements were performed on single nanostructures using a confocal microscope coupled with a laser delivering ultrafast pulses ($\Delta \tau_{pulse}$ = 160 fs, $f_{rep}$ = 80 MHz) centered at 1554 nm. The pump beam is focused using a 0.85 NA air objective, which produces a 2.2 µm beam diameter. All the experiments except the power-dependence study (**Figure 2b**) were performed with 500 µW average pump power at the sample (peak fluence 1 GW/cm$^2$). The nonlinear emission is then collected and sent either to a single-photon avalanche diode (SPAD) detector or to a spectrometer (Andor, Shamrock SR303i) equipped with a CCD camera (Andor iKon-M DU934P-BV) to acquire the emission spectrum. The power curves of SHG, THG and MPPL reported in **Figure 2b** were obtained by singling out each nonlinear emission signal by spectral filtering and detecting through the SPAD detector.

# Supplementary Material

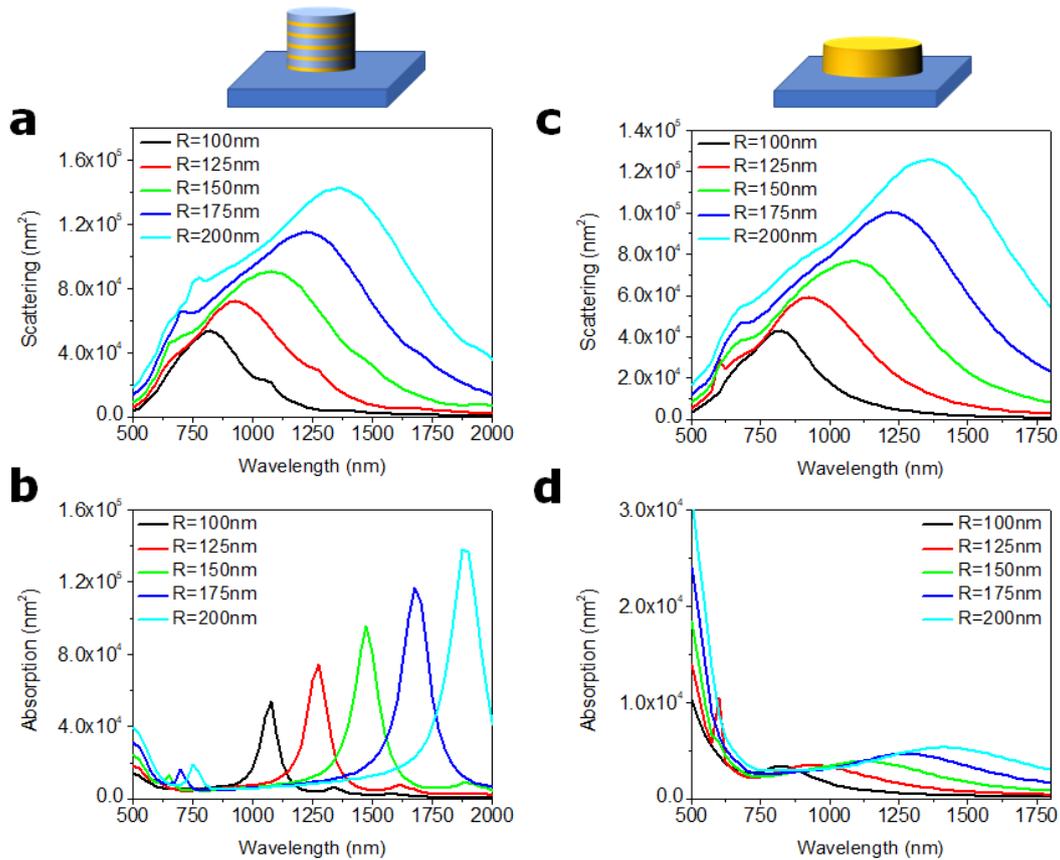

**Supplementary Figure S1.** Calculated optical (a) scattering and (b) absorption cross sections of MMD nanocavities with varying radii as a function of the exciting wavelength. Corresponding calculated optical (c) scattering and (d) absorption cross sections of bulk Au nanoresonators with the same volume of Au as in the MMD nanocavities.

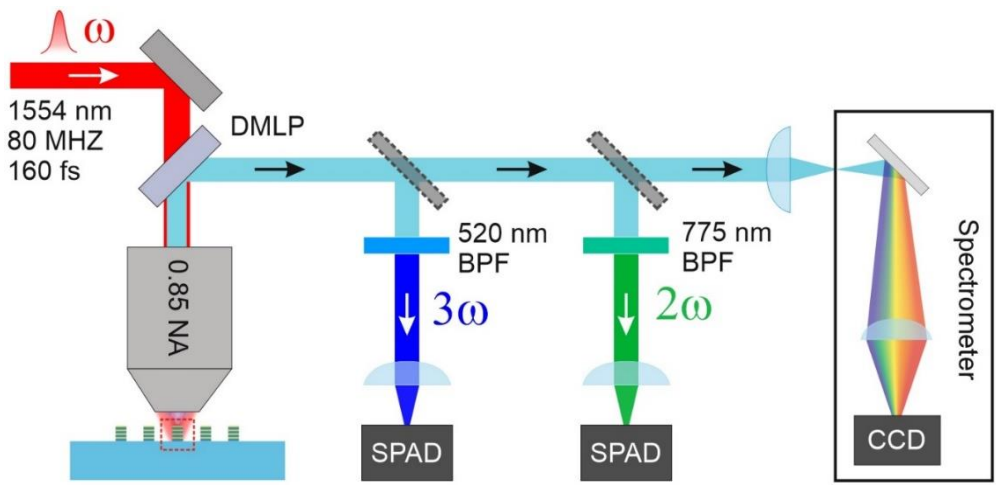

**Supplementary Figure S2.** Simplified diagram the nonlinear microscope used in the experiments. Acronyms key: DMLP = dichroic mirror, long pass; BPF = bandpass filter; NA = numerical aperture; SPAD = single-photon avalanche diode; CCD = charge-coupled device.

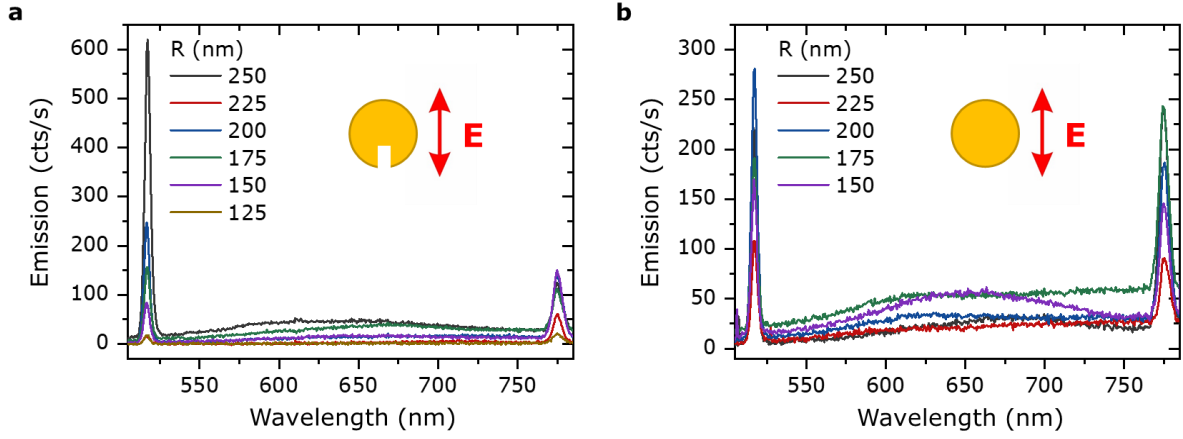

**Supplementary Figure S3.** Emission spectra of (a) multilayered "O" nanocavities and (b) multilayered "∥" nanocavities of varying radius. As we excite at 1554 nm, SHG and THG occur respectively at 517 nm and 777 nm, corresponding to the position of the two peaks visible in each panel. The broad emission band centered at 650 nm is a multiphoton photoluminescence typical of the X and L bands in gold, see Ref. **[51,52]**.

**Estimation of the SHG and THG conversion efficiencies, nonlinear coefficients and nonlinear susceptibilities.**

Let us estimate the conversion efficiencies $\eta_{\text{SHG}} = \frac{P_{\text{SHG}}^{\text{out}}}{P_{\text{FW}}^{\text{in}}}$ and $\eta_{\text{THG}} = \frac{P_{\text{THG}}^{\text{out}}}{P_{\text{FW}}^{\text{in}}}$, where $P_{\text{FW}}^{\text{in}} = 500$ µW is the average pump power used in experiments (peak fluence 1 GW/cm$^2$). We estimate as follows the optical throughput of our setup for the SHG (THG) signal. The light emitted by the nanoparticle within the solid angle of detection of the objective is 70% (70%) of the total, since we collect from the higher refractive index half-space. The light transmitted by the microscope objective is 85% (95%), while the metallic mirrors and the dichroic mirror reflects 96% (96%) and 99%, (99%) respectively. The quantum efficiency of the SPAD is 15% (45%) at 777 nm (518 nm), while the light effectively collected by the SPAD (filling factor) is 50% (80%), due to the finite size of the device (50 µm). Therefore, the overall optical throughput is 4% (20%).

Based on this estimate, the most efficient MMD antenna emits $4\times10^4/0.04 = 1\times10^6$ photons/s, which corresponds to an average power of about 300 fW at 777 nm. Hence, we obtain a conversion efficiency $\eta_{\text{SHG}} = \frac{P_{\text{SHG}}^{\text{out}}}{P_{\text{FW}}^{\text{in}}} \cong 6 \times 10^{-10}$, whereas for the bulk gold structures $\eta_{\text{SHG}}$ is of the order 10$^{-11}$. Similarly, the highest THG signal measured is $5\times10^4/0.22 = 2.3\times10^5$ photons/s, corresponding to a conversion efficiency $\eta_{\text{THG}} = \frac{P_{\text{THG}}^{\text{out}}}{P_{\text{FW}}^{\text{in}}} \cong 1.8 \times 10^{-10}$.

Some different, and perhaps more telling, figures of merit, are the nonlinear coefficients $\gamma_{\text{SHG}} = \frac{\hat{P}_{\text{SHG}}^{\text{out}}}{\left(\hat{P}_{\text{FW}}^{\text{in}}\right)^2}$ and $\gamma_{\text{THG}} = \frac{\hat{P}_{\text{THG}}^{\text{out}}}{\left(\hat{P}_{\text{FW}}^{\text{in}}\right)^3}$, obtained by dividing the peak powers ( ) of the emitted nonlinear signal $\hat{P}_{\text{SHG}}^{\text{out}}$ ($\hat{P}_{\text{THG}}^{\text{out}}$) by the square $\left(\hat{P}_{\text{FW}}^{\text{in}}\right)^2$ and the cube $\left(\hat{P}_{\text{FW}}^{\text{in}}\right)^3$ of the impinging powers respectively, where $\hat{P} = P/f_{\text{rep}}\Delta\tau_{\text{pulse}}$. In fact, these coefficients are independent from both the repetition rate, $f_{\text{rep}}$, and pulse width, $\Delta\tau_{\text{pulse}}$, of the excitation pulses. In our case these numbers are respectively $\gamma_{\text{SHG}} \cong 10^{-11} \, W^{-1}$ and $\gamma_{\text{THG}} \cong 10^{-13} \, W^{-2}$.

Finally, we shall derive the *effective* nonlinear susceptibilities $\chi_{\text{eff}}^{(2)}$ and $\chi_{\text{eff}}^{(3)}$, namely, under the simplifying assumption that all the nonlinear sources are homogeneously distributed within the nanocavity volume. Note however that such assumption might lead to a large underestimation of $\chi_{\text{eff}}^{(2)}$ since SHG occurs mostly in small interfacial volumes. In this case, following Ref. **[66]**, we obtain:

$$\chi_{\text{eff}}^{(2)} = \sqrt{\hat{\gamma}_{\text{SHG}} \frac{3\pi\epsilon_0 c A_{\text{eff}}^2}{k^4 V^2}} \cong 1.5 \text{ pm V}^{-1},$$

and

$$\chi_{\text{eff}}^{(3)} = \sqrt{\hat{\gamma}_{\text{THG}} \frac{1.5\pi\epsilon_0^2 c^2 A_{\text{eff}}^3}{k^4 V^2}} \cong 1.1 \times 10^{-20} \text{ m}^2 \text{ V}^{-2}.$$

where $A_{\text{eff}} = \pi \times (0.67 \times 1554 \text{ nm}/0.85)^2 = 4.7 \times 10^{-12} \text{ m}^2$ is the area of the diffraction-limited pump and $V = \pi \times (200 \text{ nm})^2 \times 400 \text{ nm} = 5.0 \times 10^{-20} \text{ m}^3$ is the volume of the most resonant structures.

**Author Contribution**

NM and MC conceived the idea. NM designed the structures and performed the numerical calculations with inputs from AZ, MF and MC. AZ, LG and MC designed the experiments and performed the nonlinear optical response characterization. NM, AZ, MF and MC analyzed the data, interpreted the results and wrote the manuscript. TI, MI and FDA fabricated and imaged the samples, and participated in the general discussion. NM and AZ contributed equally.

**Acknowledgements**

NM acknowledges support from the Luxembourg National Research Fund (Grant No. C19/MS/13624497 'ULTRON') and from the FEDER Program (Grant No. 2017-03-022-19 'Lux-Ultra-Fast'). The present work has been carried out in the framework of the PRIN 2017 project "NOMEN" (Grant n. 2017MP7F8F) funded by the Italian Ministry of University and Research (MIUR).